\documentclass[a4paper,10pt]{article}
\usepackage{enumerate}
\usepackage{color}
\usepackage[utf8]{inputenc} 
\usepackage[english]{babel}
\usepackage[T1]{fontenc}
\usepackage{graphicx}
\usepackage{amsfonts,amssymb,amsmath,latexsym,amsthm}
\usepackage{textcomp}
\usepackage[pdftex]{hyperref}
\usepackage{geometry}
\title{Dynamics of irregular wave fields in the Schamel equation framework}
\author{Marcelo V. Flamarion$^{1}$, Efim Pelinovsky$^{2,3}$ and Ekaterina Didenkulova$^{2,3}$}
\date{}

\begin{document}
\maketitle
\begin{center}
{\footnotesize $^1$Departamento Ciencias—Secci{\' o}n Matem{\' a}ticas, Pontificia Universidad Cat{\' o}lica del Per{\' u}, Av. Universitaria 1801, San Miguel 15088, Lima, Peru  \\
mvellosoflamarionvasconcellos@pucp.edu.pe}

\vspace{0.3cm}
{\footnotesize 
$^{2}$Faculty of Informatics, Mathematics and Computer Science, HSE University, Nizhny Novgorod 603155, Russia.

$^{3}$ Gaponov-Grekhov Institute of Applied Physics, Nizhny Novgorod, 603122, Russia.}



\end{center}


\begin{abstract} 
The present article is devoted to the study of the dynamics of narrowband wave fields within the non-integrable Schamel equation, which plays an important role in plasma physics, wave dynamics in metamaterials, and electrical circuits. A Monte Carlo approach is used to obtain a large number of random independent realizations of the wave fields, allowing for an investigation of the evolution of the following statistical characteristics: spectra, moments, and distribution functions. The simulations are conducted for different values of the Ursell number (the ratio of nonlinearity to dispersion) to study the impact of nonlinearity and dispersion on the processes under consideration.

	\end{abstract}

\section{Introduction}
Research often focuses on equations that can be integrated using the inverse scattering method. These equations possess an infinite number of conserved integrals (invariants), and many solution parameters can be obtained analytically. However, when non-analytical forms of nonlinearity are introduced, the equation’s integrability is disrupted. This disruption can result in inelastic wave collisions and the emergence of additional radiation due to wave interactions. In such cases, analytical methods become inadequate, and numerical modeling becomes the primary tool for studying wave field dynamics. Occasionally, the non-integrability of an equation may have only a minor effect on its solutions. For example, in \cite{Dutykh2014}, it was shown that the interaction of solitons with the same polarity within the integrable Korteweg–de Vries equation and the non-integrable Benjamin-Bona-Mahony equation is remarkably similar, with no significant changes in wave field characteristics during the evolution process.

A similar conclusion was drawn when comparing the interaction of unipolar solitons in the KdV equation with those in the non-integrable Schamel equation \cite{Flamarion2023}. However, in the case of solitons with opposite polarities, the non-integrability of the equation can lead to effects not observed in integrable models. For instance, in the studies by Flamarion et al. \cite{Flamarion2024} and Didenkulova et al. \cite{Didenkulova2023}, the non-integrable Schamel model demonstrated energy transfer from a smaller soliton to a larger one during interaction, which can ultimately result in the formation of an anomalously large wave through multiple wave collisions. This phenomenon has also been noted in the non-integrable variant of the nonlinear Schrödinger equation \cite{Krylov1980,Zakharov1988,Dyachenko1989,Zakharov2012}. Such effects are impossible in integrable equations, where soliton interactions are elastic, as seen in the modified KdV and Gardner equations \cite{Ruderman:2008, Ruderman:2023}. Therefore, non-integrable models are of significant interest for further research.

The Schamel equation was originally derived in plasma physics \cite{Schamel:1972, Schamel1973}, and in recent years, it has been actively used to describe waves in metamaterials \cite{Zemlyanukhin2019, Mogilevich2023}, electrical circuits \cite{Kengne2020, Aziz2020}, damping systems \cite{Shan:2019, Sultana:2022}, and other problems in plasma physics \cite{Williams:2014, Ali:2017, Chowdhury:2018, Mushtaq:2006, Saha:2015a, Saha:2015b}. Unlike the Korteweg-de Vries equation, the nonlinear term of the Schamel equation contains the modulus of the wave function. Since the Schamel equation is non-integrable, analytical methods are not applicable for analyzing wave dynamics, necessitating numerical integration. Within the framework of this equation, three invariants are preserved: mass, momentum (energy), and the Hamiltonian \cite{Flamarion2023}. These invariants play a decisive role in assessing the accuracy and reliability of numerical solutions to the equation. In this paper, we study the dynamics of irregular wave fields (wave turbulence) within the framework of the Schamel equation and determine the effect of nonlinearity on the statistical characteristics of wave fields. Similar problems within the framework of integrable equations were considered earlier: \cite{Pelinovsky2006, Didenkulova2019} within the KdV framework, and \cite{Flamarion2023a} within the Benjamin–Ono and Boussinesq equations \cite{Flamarion2024b}. The effect of the non-integrability of the model on the characteristics of wave turbulence remains to be clarified.

The article is organized as follows: In Section 2  we introduce the Schamel equation and the numerical methods. In Section 3  we present the results, followed by the final conclusions in Section 4.

\section{The Schamel equation}
In our research, we investigate wave field dynamics within the Schamel equation in the form
\begin{equation}\label{Schamel1}
u_{t} +\sqrt{|u|}u_{x}+\frac{1}{U_{r}}u_{xxx}=0.
\end{equation}
The variable $u$ represents the wave field as the function of horizontal coordinate and time
Ursell number ($U_r$) specifies the ratio of nonlinearity to dispersion parameter \cite{Karpman1973}. When is small, the dispersive term dominates, otherwise the
nonlinearity plays the dominant role.

To model the initial wave field, we consider an irregular wave field represented by a Fourier series with $N = 256$ harmonics
\begin{equation}\label{initial}
u(x,0) = \sum_{i=1}^{M}\sqrt{2S(k_i)\Delta k}\cos(k_{i}x+\varphi_{i}).
\end{equation}
In this context, $S(k)$ stands for the initial power spectrum, where $k_i = i\Delta k$, with $\Delta k$ as the sampling wavenumber, and $\varphi_i$ is a random variable uniformly distributed over the interval $(0, 2\pi)$. The length of the initial realization is given by $L = 2\pi/\Delta k$. In this study, we will assume a Gaussian-shaped wavenumber power spectrum.
\begin{equation}\label{Gaussian}
S(k) = Q\exp\Big(-\frac{1}{2}\frac{(k-k_0)^2}{2K^2}\Big), \;\ k>0.
\end{equation}
For numerical simulation the following meanings of the parameters set: the dimensionless peak wavenumber $k_0=1$, the spectral width $K=0.18$. The relative energy $Q$ is chosen so that
the wave field has the same total energy, which is defined by the variance $\sigma^2 = 0.25$
\begin{equation}\label{sigma}
\sigma^{2}=\frac{1}{L}\int_{0}^{L}u^{2}(x,t)dx.
\end{equation}

The evolution of the wave field is computed using the classical pseudospectral method outlined in \cite{Trefethen2000}. We solve the equation within a periodic computational domain of $[0, 2\pi / \Delta k]$, where $\Delta k = 0.023$. To ensure a precise approximation of spatial derivatives, a uniform grid with $N = 2^{12}$ points is utilized \cite{Trefethen2000}. For time integration, we employ the fourth-order Runge-Kutta method with discrete time steps of size $\Delta t = 0.005$. This approach has been successfully applied in similar numerical simulations involving the Schamel equation across various contexts \cite{Flamarion2023, Flamarion2024, Flamarion2023a, Flamarion2023b} and the KdV equation \cite{Pelinovsky2006, Flamarion2021}.

\section{Results}
\begin{figure}[h!]
\centering	
\includegraphics[scale=1]{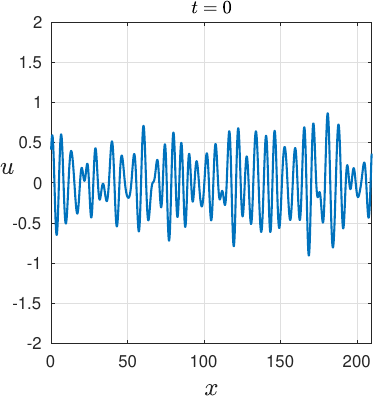}
\includegraphics[scale=1]{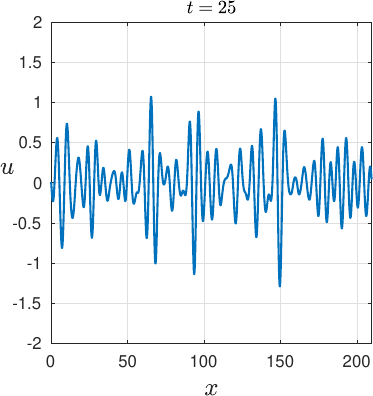}
\includegraphics[scale=1]{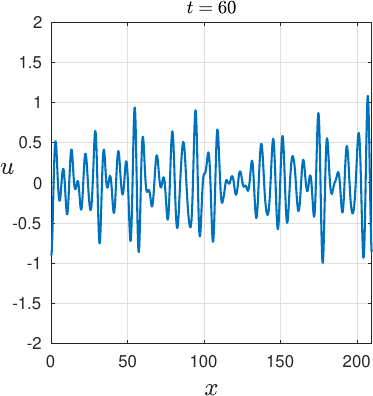}
\includegraphics[scale=1]{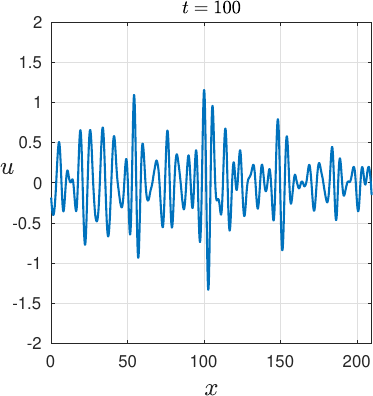}
\caption{ The wave field at different times with $U_r=1$.} 
\label{Fig1}
\end{figure}

\begin{figure}[h!]
\centering	
\includegraphics[scale=1]{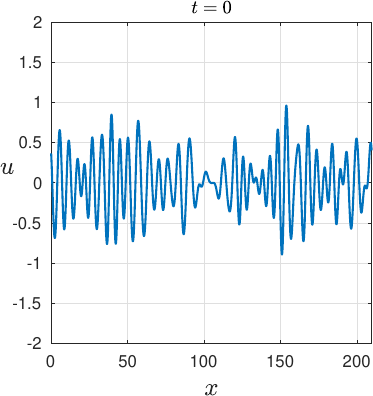}
\includegraphics[scale=1]{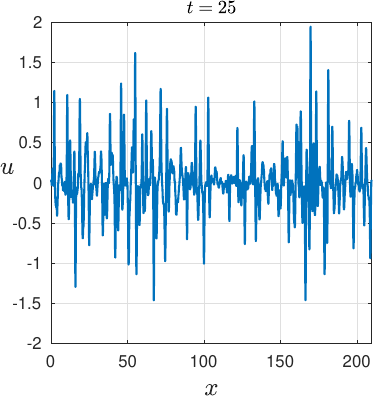}
\includegraphics[scale=1]{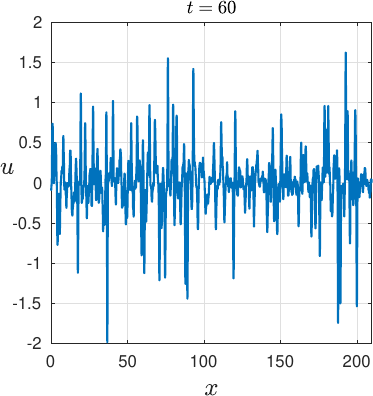}
\includegraphics[scale=1]{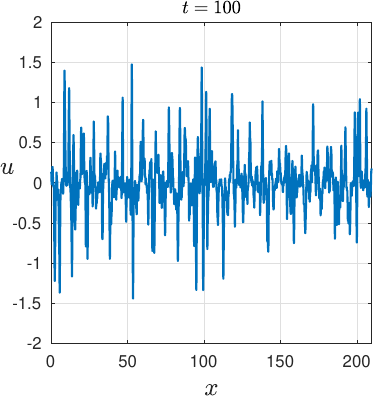}
\caption{ The wave field at different times with $U_r=100$.} 
\label{Fig2}
\end{figure}

The wave field is Gaussian and symmetric at the initial moment of time. The evolution of the wave fields for $U_r=1$ and $U_r=100$ is shown in Figure \ref{Fig1} and Figure \ref{Fig2}, respectively. During the evolution of the wave field with small $U_r$, individual waves with large heights are formed, which stand out against the background of the other waves. However, in the case of a wave field with 
$U_r=100$, this effect is much stronger. Individual waves become very sharp, and the heights increase significantly due to high nonlinearity. The wave envelope ceases to be smooth, and the signal is narrowband.

\begin{figure}[h!]
\centering	
\includegraphics[scale=1]{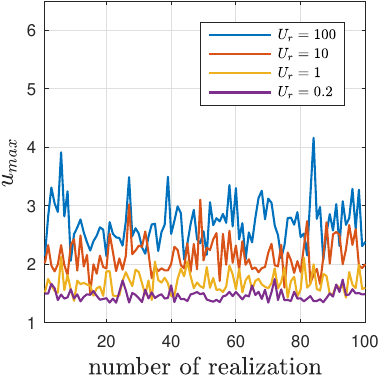}
\includegraphics[scale=1]{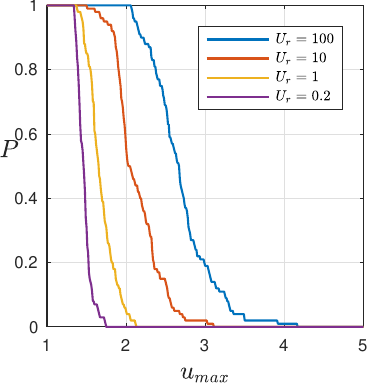}
\caption{ (Left): Maximum of wave amplitudes over time. (Right): Distribution of maximum crest amplitudes over 100 realizations for different values of $U_r$.} 
\label{Fig3}
\end{figure}

Increasing the nonlinearity of the wave field leads to higher wave heights. This is clearly illustrated in Figure \ref{Fig3}, which shows the maximum values of the wave fields and the wave amplitude distribution functions for different values of the similarity parameter. Greater nonlinearity results in higher maximum values of the wave fields. It should also be noted that the fluctuations of the curves in the left graph of Figure \ref{Fig3} increase with the increasing Ursell number.

\begin{figure}[h!]
\centering	
\includegraphics[scale=1]{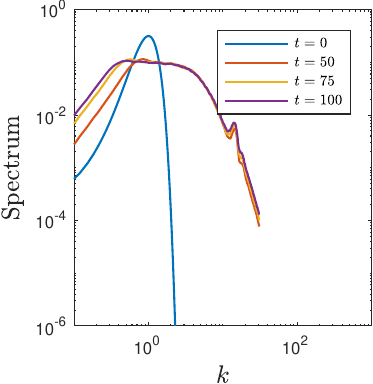}
\includegraphics[scale=1]{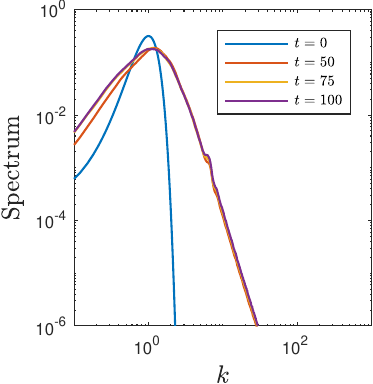}
\includegraphics[scale=1]{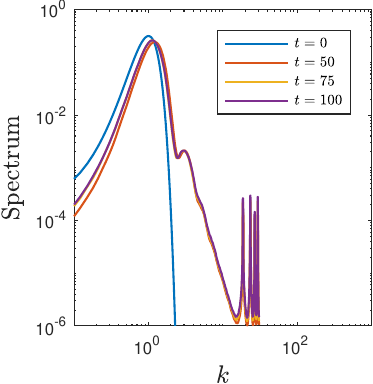}
\includegraphics[scale=1]{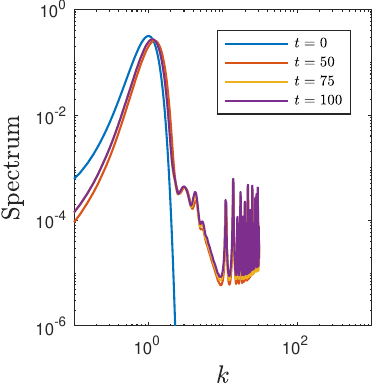}
\caption{ The averaged of spectra over 100 realization of the wave fields at different times. From (top) to (bottom) and (left) to (right) $U_r =100$, $U_r =10$, $U_r =1$ and $U_r =0.2$.} 
\label{Fig4}
\end{figure}

The evolution of the wave spectra, averaged over 100 realizations for different degrees of nonlinearity in the wave fields, is shown in Figure \ref{Fig4}. In all cases, the spectra tend to reach a stationary state as they evolve over time. For strongly nonlinear wave fields with a large similarity parameter, the spectra expand over time, and their peak becomes flattened. This indicates a spectrum downshift, which redistributes energy towards the region of smaller wavenumbers. This effect is consistent with the behavior observed in wave spectra of integrable models \cite{Pelinovsky2006, Flamarion2023}. During the evolution, a relatively strong high-frequency tail emerges in the spectrum (at the level of $10^{-4}$), particularly for small Ursell numbers (Figure \ref{Fig4}). This phenomenon may be attributed to the non-analytic nature of the nonlinear function in the Schamel equation. At small Ursell numbers, the generation of the high-frequency spectrum can be described within the framework of perturbation theory, representing the wave field in the form
\begin{equation}
u(x,t) = A(t)\cos(\omega t-kx)+v(x,t),
\end{equation}
where $v$ is a small correction to the main wave. In the first order of perturbation theory, we have the following equation for it:
\begin{equation}
\frac{\partial v}{\partial t} + \frac{1}{U_r^3} v_x^3 = -\frac{2}{3} \frac{\partial}{\partial x} A^{3/2} \cos^{3/2} (\omega t - kx).
\end{equation}
This equation is linear, and the "external" force (the right-hand side) contains many harmonics. (If the nonlinearity were quadratic, the right-hand side would contain only the second harmonic with a constant amplitude). Accordingly, they also generate many harmonics in the correction $v(x,t)$. Their amplitudes are small, and they interact poorly with the main wave, remaining noticeable over long times. If the nonlinearity is large, there is a strong process of energy transfer between different harmonics, leading to a continuous spectrum, as shown in Figure \ref{Fig4}.

Let us consider the behavior of higher statistical moment
\begin{equation}\label{moments}
\mu_{n}(t)=\frac{1}{L}\int_{0}^{L}u^{n}(x,t)dx, \mbox{ where $n=1,2,3,4.$}
\end{equation}
More specifically, we concentrate on two statistical quantities that characterize the wave spectrum: the kurtosis excess ($\kappa$) and the skewness ($\varsigma$), which are defined as
\begin{equation}\label{kurtskew}
\kappa(t) = \frac{\mu_{4}}{\mu_{2}^{2}}-3 \mbox{ and } \varsigma(t) = \frac{\mu_{3}}{\mu_{2}^{3/2}}.
\end{equation}

\begin{figure}[h!]
\centering	
\includegraphics[scale=1]{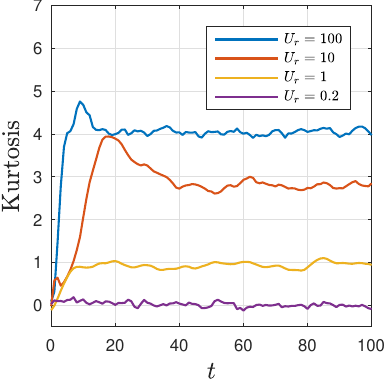}
\includegraphics[scale=1]{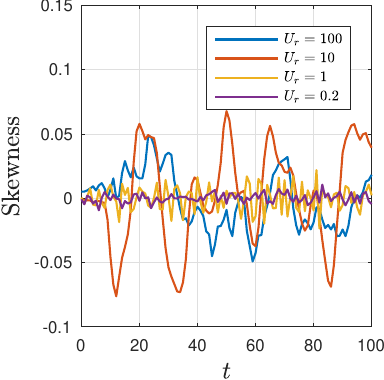}
\caption{Temporal evolution of the skewness and kurtosis for different values of the parameter $U_r$ averaged over 100 realizations.} 
\label{Fig5}
\end{figure}

The evolution of these quantities (see Fig.~\ref{Fig5}) fully corresponds to the conclusions made above. It is evident that for small values of $U_r$, averaged over 100 realizations, the skewness is close to zero and experiences small fluctuations. However, the amplitude and period of skewness deviations are significantly higher for large $U_r$, indicating a strong change in the behavior of the wave field and the predominance of positive and negative waves at different moments in time. At $U_r = 0.2$, the average kurtosis is close to zero, demonstrating proximity to a Gaussian process and a low probability of the occurrence of large waves. In addition, the higher the value of the Ursell number, the more nonlinear the wave field becomes, and the greater the kurtosis. In all the cases considered, the average kurtosis reaches a stationary state. A short-term transient period of strong kurtosis change is observed, which was previously noted for the evolution of random fields and soliton ensembles \cite{Flamarion2023,Didenkulova2019}. The behavior of moments resembles similar scenarios for deep water and intermediate depths, as observed in studies based on the nonlinear Schrödinger equation \cite{Tanaka2001,Onorato2001}.

\begin{figure}[h!]
\centering	
\includegraphics[scale=1]{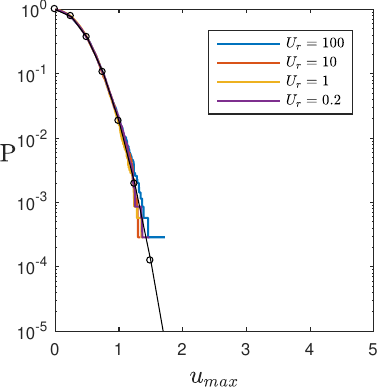}
\includegraphics[scale=1]{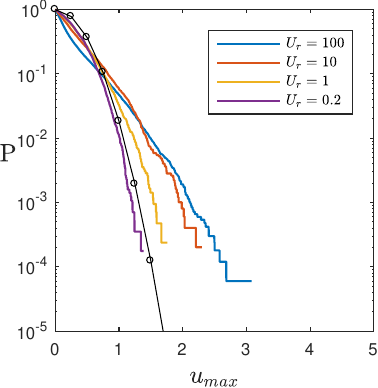}
\caption{The exceedance probability distributions for wave crests at the initial time $(t = 0)$ and at the asymptotic state $(t = 100)$ (right) are displayed for various $U_r$ values across 100 realizations. The black solid line with circles indicates the Rayleigh distribution associated with a narrow-band Gaussian process.
} 
\label{Fig6}
\end{figure}

\begin{figure}[h!]
\centering	
\includegraphics[scale=1]{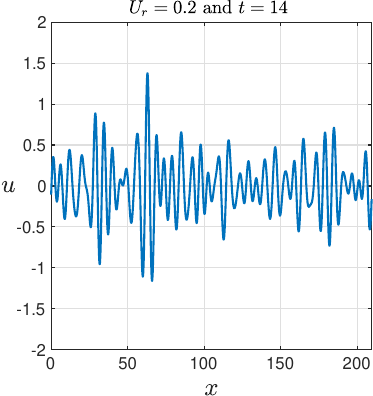}
\includegraphics[scale=1]{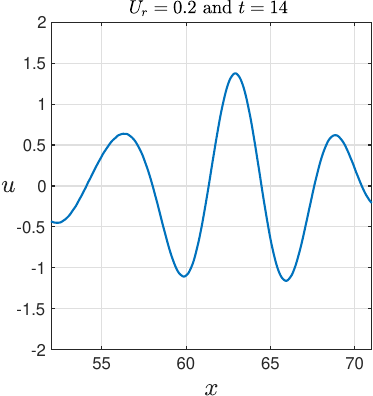}
\includegraphics[scale=1]{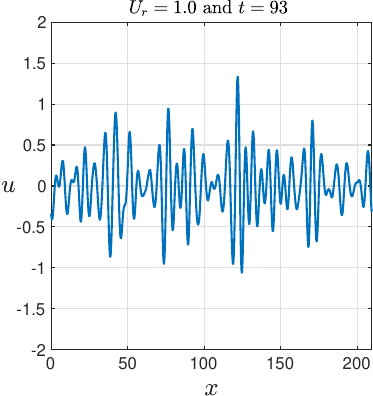}
\includegraphics[scale=1]{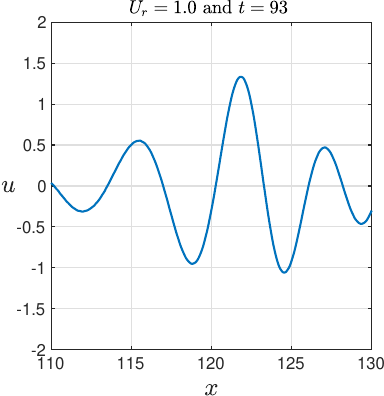}
\caption{Wave fields containing rogue waves (left) and rogue wave profiles (right).} 
\label{Fig7}
\end{figure}

The wave crest distribution functions for wave fields with different values of the $U_r$ number are shown in Figure \ref{Fig6}. At the initial time ($t = 0$), the distributions, as expected, coincide fairly well with the Rayleigh distribution. The only exception is the region of large waves, where the curves for $U_r$ less than 10 lie below the Rayleigh curve, and for $U_r = 100$, they lie above the Rayleigh curve. At large times, the curves shift significantly. Only the curve corresponding to the smallest $U_r$ still lies below the Rayleigh curve, as in the initial distribution. This also correlates with the kurtosis, which is close to zero and even negative at some times for this $U_r$ number. The tails of the other functions shift significantly to the region of large amplitudes, where the most extreme values correspond to the largest $U_r$.
It follows from the above analysis that the evolution of wave fields leads to the emergence of waves with abnormally large amplitudes (rogue waves). Wave profiles (and the corresponding wave fields) with maximum amplitudes observed in numerical experiments for all considered values of the Ursell number are shown in Fig.~\ref{Fig7}.

\begin{figure}[h!]
\centering	
\includegraphics[scale=1]{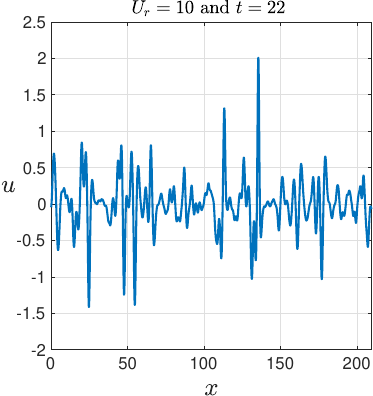}
\includegraphics[scale=1]{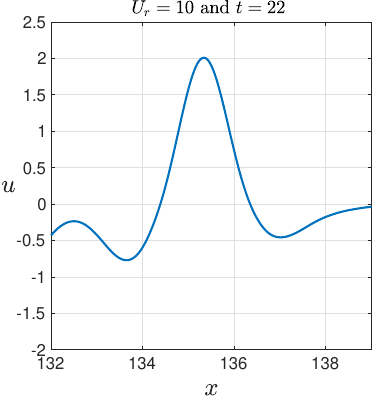}
\includegraphics[scale=1]{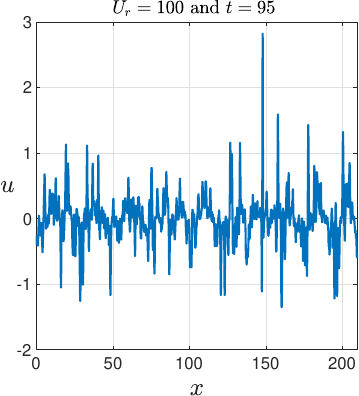}
\includegraphics[scale=1]{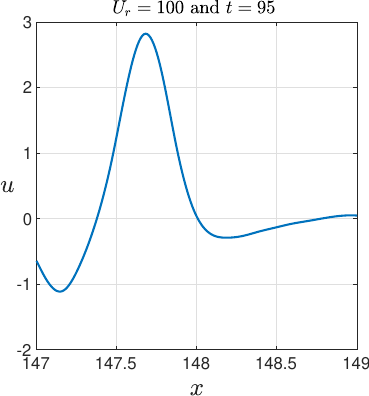}
\caption{Wave fields containing rogue waves (left) and rogue wave profiles (right).} 
\label{Fig8}
\end{figure}

In oceanology, a rogue wave is defined as a wave whose height $H_{\text{fr}}$ is at least twice the significant wave height $H_s$:
\[
\frac{H_{\text{fr}}}{H_s} > 2. \tag{9}
\]
The significant wave height is defined as the average of 1/3 of the highest waves in the record. The largest freak waves, as calculated, are shown in Figures \ref{Fig7} and \ref{Fig8}. The wave fields containing freak waves are given on the left, and the portraits of freak waves are shown on the right. The abnormally large waves have the following parameters: $ H_{\text{fr}}/H_s = 2 $ in the case of $ Ur = 0.2 $; $ H_{\text{fr}}/H_s= 2 $ in the case of $ Ur = 1 $; $ H_{\text{fr}}/H_s= 2.7 $ in the case of $ Ur = 10 $; and $ H_{\text{fr}}/H_s= 4.3 $ for $ Ur = 100 $.

\section{Conclusion}
The statistical analysis of the irregular wave fields is conducted numerically within the framework of the non-integrable Schamel equation. The main focus is on examining the influence of the nonlinearity rate (Ursell number) on the statistical characteristics of the wave ensembles. For large values of the Ursell number, the spectra expand but remain quasi-symmetric. In contrast, for wave fields with relatively small nonlinearity, the spectrum narrows in the region of small wavenumbers, highlighting the relative symmetry of the wave fields during evolution. An increase in the nonlinearity of the wave field leads to greater extremeness, meaning a higher probability of observing anomalously large waves. As the Ursell number increases, the kurtosis rises significantly, along with the tails of the wave amplitude distribution functions. This observation is consistent with similar findings for integrable systems. For wave fields with small $U_r$, the amplitude distribution function remains below Rayleigh for large times, and the kurtosis fluctuates around zero, periodically taking negative values. The emergence of freak waves in wave fields with different Ursell numbers is also demonstrated.

\section{Acknowledgements}
E.P. and E.D. were supported by the Russian Science Foundation, Grant No. 22-17-00153.

	\section*{Declarations}
	
	\subsection*{Conflict of interest}
	The authors state that there is no conflict of interest. 
	\subsection*{Data availability}
	
	Data sharing is not applicable to this article as all parameters used in the numerical experiments are informed in this paper.

\end{document}